\documentclass[12pt,prd,showpacs,preprint,amsmath,amssymb]{revtex4}
\usepackage{graphicx}
\usepackage{epsfig}
\usepackage{dcolumn}
\usepackage{bm}
\include{def-com}

\newcommand{\jpsi}{J/\psi}
\newcommand{\psip}{\psi(2S)}

\begin{document}
\title{ \boldmath First observation of  $\psip \to p \bar{n} \pi^- +c.c.$}
\author{
M.~Ablikim$^{1}$, J.~Z.~Bai$^{1}$, Y.~Ban$^{12}$,
J.~G.~Bian$^{1}$, X.~Cai$^{1}$, H.~F.~Chen$^{17}$,
H.~S.~Chen$^{1}$, H.~X.~Chen$^{1}$, J.~C.~Chen$^{1}$,
Jin~Chen$^{1}$
% ...
\\(BES Collaboration)\\
\vspace{0.2cm} {\it
$^{1}$ Institute of High Energy Physics, Beijing 100049, People's Republic of China\\
% ...
$^{12}$ Peking University, Beijing 100871, People's Republic of China\\
% ...
$^{17}$ University of Science and Technology of China, Hefei 230026, People's Republic of China\\
% ...
} 
}

\begin{abstract}

Using 14 million $\psi(2S)$ events collected with the Beijing
Spectrometer (BESII) at the Beijing Electron-Positron Collider, the
branching fractions of $\psip$ decays to $p \bar{n} \pi^-$ and
$\bar{p} n \pi^+$ and the branching fractions of the main background
channels $\psip \to p \bar{n} \pi^-\pi^0$, $\psip \to \gamma\chi_{c0}
\to \gamma p \bar{n} \pi^-$, $\psip \to \gamma\chi_{c2} \to \gamma p
\bar{n} \pi^-$, and $\psip \to \gamma \chi_{cJ} \to \gamma p \bar{n}
\pi^- \pi^0$ are determined. The contributions of the $N^{\ast}$
resonances in $\psip \to p \bar{n} \pi^- +c.c.$ are also discussed.

\end{abstract}

\pacs{14.20.Gk, 13.75.Gx, 13.25.Gv}

\maketitle

\section{\boldmath Introduction}

From perturbative QCD (pQCD), it is expected that both $J/\psi$ and
$\psip$ decaying into light hadrons are dominated by the annihilation
of $c \bar c$ into three gluons, with widths proportional to the
square of the wave function at the origin
$|\Psi(0)|^2$~\cite{T.Appelquist}. This yields the pQCD ``$12\%$
rule''
 \[ Q_h = \frac{B_{\psip\to h}}{B_{J/\psi\to h}} \approx
 \frac{B_{\psip\to e^+e^-}}{B_{J/\psi\to e^+e^-}} \approx 12\% .\] The
violation of this rule was first observed in the $\rho \pi$ and
$K^{\ast +}K^- +c.c.$ decay modes by Mark-II~\cite{Mark-II}. Following
the scenario proposed in Ref.~\cite{J.L.Rosner}, that the small $\psip
\to \rho \pi$ branching fraction is due to the cancellation of the
$S$- and $D$-wave matrix elements in $\psip$ decays, it was suggested
that all $\psip$ decay channels should be affected by the same $S$-
and $D$-wave mixing scheme, and thus all ratios of branching fractions
of $\psip$ and $J/\psi$ decays into the same final state could have
values different from $12\%$, expected between pure $1S$ and $2S$
states~\cite{P.Wang}. The mixing scenario also predicts $\psi(3770)$
decay branching fractions since the $\psi(3770)$ is a mixture of $S$-
and $D$-wave charmonia, as well.  Many channels of $J/\psi$, $\psip$,
and $\psi(3770)$ decays should be measured to test this scenario.

A very important source of information on nucleon internal structure
is the $N^{\ast}$ mass spectrum, including production and decay
rates. Because of its importance for the understanding of
nonperturbative QCD, a series of experiments on $N^{\ast}$ physics
with electromagnetic probes (real photons and electrons with
space-like virtual photons) are being performed at facilities such
as JLAB, ELSA at Bonn, GRAAL at Grenoble, and SPRING8 at
JASRI~\cite{pro}. They have already produced some
results~\cite{M.Ripani,T.Nakano}. However, our knowledge on
$N^{\ast}$ resonances is still poor. Even for the well-established
lowest excited state, the $N^{\ast}(1440)$, properties such as mass,
width, and decay branching fractions still have large experimental
uncertainties~\cite{pdg}.  Another outstanding problem is that, in
many of its forms, the quark model predicts a substantial number of
$N^{\ast}$ states around 2 $\hbox{GeV}/c^2$, which have not yet been
observed~\cite{S.Capstick}.

Recent studies of $N^{\ast}$ resonances have been performed using
$\jpsi$ events collected at the Beijing Electron-Positron Collider
(BEPC)~\cite{zou,jxb}, providing a new method for probing
this physics, and a new $N^{\ast}$ peak with a mass at around
2065 $\hbox{MeV}/c^2$ was observed~\cite{jxb}. This may be one of the
``missing" $N^{\ast}$ states around 2~$\hbox{GeV}/c^2$. However, due
to its large mass, the production of this $N^{\ast}(2065)$ in $\jpsi$
decays is rather limited in phase space. A similar search for it
in $\psip\to p\bar{p} \pi^0$ has been performed~\cite{xinbo}, where
there is a faint but not statistically significant accumulation of
events in the $p \pi$ invariant mass spectrum at around 2065
$\hbox{MeV}/c^2$.

In this paper, we study $\psip\to p \bar{n} \pi^- +c.c.$ and their
main background channels, determine branching fractions, test the
$12\%$ rule, and study $N^{\ast}$ resonances in the $N
\pi$ system.

\section{\boldmath BES detector and the data sample}

BESII is a large solid-angle magnetic spectrometer which is
described in detail in Ref.~\cite{detector}. The momentum of
charged particles is determined by a 40-layer cylindrical main
drift chamber (MDC) which has a momentum resolution of
$\sigma_{p}$/p=$1.7\%\sqrt{1+p^2}$ ($p$ in $\hbox{GeV}/c$).
Particle identification (PID) is accomplished using specific
ionization ($dE/dx$) measurements in the drift chamber and
time-of-flight (TOF) information in a barrel-like array of 48
scintillation counters. The $dE/dx$ resolution is
$\sigma_{dE/dx}\simeq8.0\%$; the TOF resolution for Bhabha events
is $\sigma_{TOF}= 180$ ps.  Radially outside of the time-of-flight
counters is a 12-radiation-length barrel shower counter (BSC)
comprised of gas tubes interleaved with lead sheets. The BSC
measures the energy and direction of photons with resolutions of
$\sigma_{E}/E\simeq21\%/\sqrt{E}$ ($E$ in GeV),
$\sigma_{\phi}=7.9$ mrad, and $\sigma_{z}=2.3$ cm. The iron flux
return of the magnet is instrumented with three double layers of
proportional counters (MUC) that are used to identify muons.

In the analysis, a GEANT3-based Monte Carlo (MC) simulation program
(SIMBES) with detailed consideration of the detector performance is
used. The consistency between data and Monte Carlo has been checked in
many high purity physics channels, and the agreement is
reasonable~\cite{simbes}. For the MC generators, the angular distribution
for $\psip \to \gamma \chi_{cJ}$ is simulated assuming a pure $E1$
transition, and uniform phase space is used for the other decays.

The data sample used for this analysis consists of $(14.0\pm 0.6)
\times 10^6$ $\psip$ events taken at $\sqrt s
=3.686$~GeV~\cite{X.H.Mo}. Backgrounds are estimated using an
inclusive $\psip$ decay MC sample generated by LUNDCRM~\cite{chenjc}
with the same size as the $\psip$ data.

\section{\boldmath Event Selection}

For the signal channel $\psip \to p \bar{n} \pi^- +c.c.$ and the
background channels, we reconstruct two charged tracks, and the
neutron and antineutron are not measured.  However, since most
antineutrons annihilate in the detector (mainly in the BSC) and most
neutrons pass through the detector without interaction, these
signatures are used to suppress backgrounds by requiring a neutral
cluster in the expected antineutron direction and requiring nothing in
the neutron direction.

A neutral cluster is considered to be a good photon candidate if the
following requirements are satisfied: it is located within the BSC
fiducial region; the energy deposited in the BSC is greater than 50
MeV; the first hit appears in the first 6 radiation lengths;  the
angle between the cluster development direction in the BSC and the
photon emission direction from the beam interaction point (IP) is
less than $37^\circ$; and the angle between the cluster and the
nearest charged particle is greater than $15^\circ$.

   Each charged track is required to be well fit by a three
dimensional helix, to originate from the interaction point,
$V_{xy}=\sqrt{V_x^2+V_y^2}<1$ cm, $|V_z|<15$ cm, and to have a polar
angle $|\cos\theta|<0.8$. Here $V_x$, $V_y$, and $V_z$ are the $x$,
$y$, and $z$ coordinates of the point of closest approach to the
beam axis. The TOF and $dE/dx$ measurements for each charged track
are used to calculate $\chi_{PID}^2(i)$ values and the corresponding
confidence levels $Prob_{PID}(i)$ for the hypotheses that a track is
a pion, kaon, or proton, where $i~(i=\pi/K/p)$ is the particle type.

We use the following common selection criteria for
all channels:
\begin{enumerate}
  \item The number of charged tracks in the MDC is two with net charge
  zero, and the difference in $V_z$ between the positive and
  negative charged tracks $|V_z^+ -V_z^-|$ is required to be less
  than 3 cm.
  \item For each charged track, the  particle
  identification confidence level for
  a candidate particle assignment is required to be greater than
  0.01. For a proton we also require $Prob_{PID}(p)>Prob_{PID}(\pi)$ and
  $Prob_{PID}(p)>Prob_{PID}(K)$, and
  for a $\pi$, we require $Prob_{PID}(\pi)>Prob_{PID}(p)$
  and $Prob_{PID}(\pi)>Prob_{PID}(K)$.
  \item The energy of the positive charged track observed in the BSC
        should be less than 0.7 GeV in order to
        eliminate $\psip\to e^+e^-$ and Bhabha events.
  \item $M_{p\pi^- (\hbox{or}\  \bar{p} \pi^+)}>1.15$ $\hbox{GeV}/c^2$
   to remove background channels with $\Lambda$ or $\bar{\Lambda}$.
 \end{enumerate}

\begin{figure}[htbp]
\epsfig{file=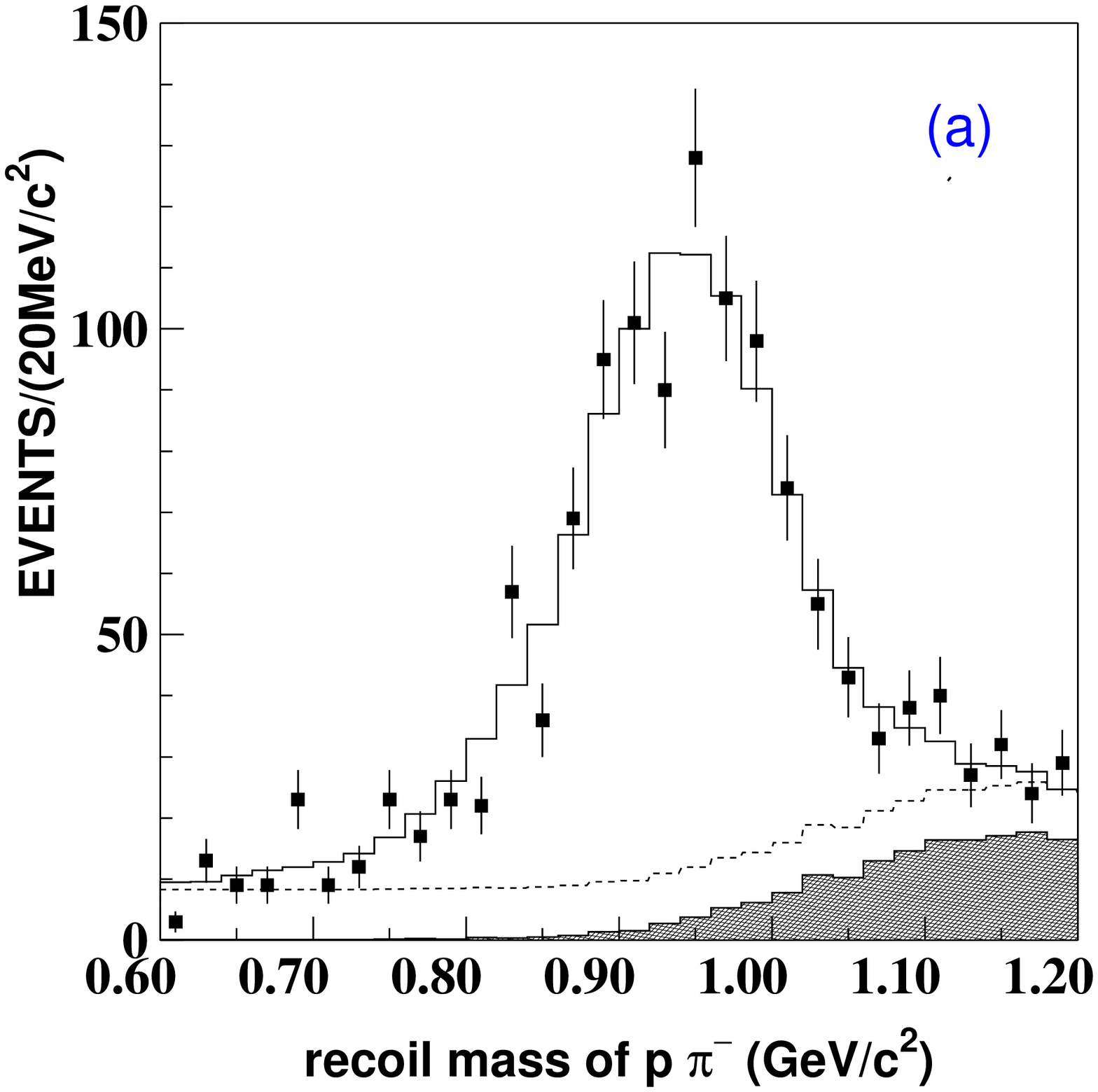,width=6.0cm}
\epsfig{file=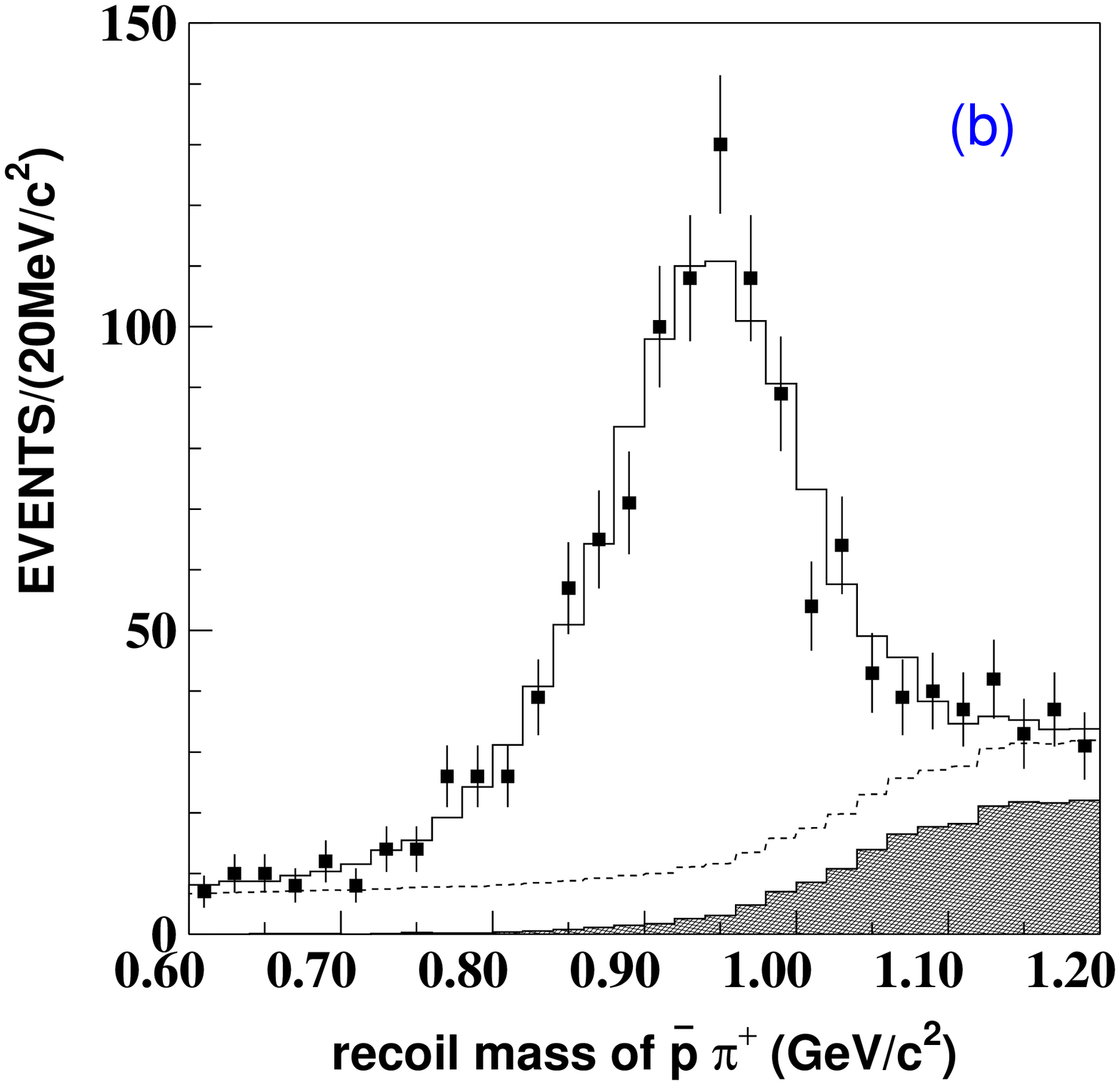,width=6.0cm} \caption{\label{pnpi-data}
Missing mass distributions of (a) $p \pi^-$ and (b) $\bar{p} \pi^+$
 of $\psip \to p \bar{n} \pi^- +c.c.$ candidate
events. The squares with error bars are data, the blank histograms
are the fit, the shaded histograms are normalized backgrounds from
$\psip \to \gamma \chi_{cJ}, \chi_{cJ} \to p \bar{n} \pi^- +c.c.$,
$\psip \to p \bar{n} \pi^- \pi^0+c.c.$, $\psip \to \gamma \chi_{cJ},
\chi_{cJ} \to p \bar{n} \pi^- \pi^0+c.c.$, and the dashed curves are
background shapes from the fit.}
\end{figure}

For $\psip \to p \bar{n} \pi^- +c.c.$ when there are photon
candidates in the events, $\alpha<10^{\circ}$ is required, where
$\alpha$ is the angle between the nearest neutral cluster and the
missing momentum direction of the two charged tracks. When there are
no photon candidates in an event, the value of $\alpha$ is set to
zero degree. After all above requirements, we obtain the $p \pi^-$
and $\bar{p} \pi^+$ missing mass distributions shown in
Fig.~\ref{pnpi-data}, where there are clear signals of $n$ and
$\bar{n}$ at around 0.94 $ \hbox{GeV}/c^2$. The background level
increases from about 10 events per 20 $\hbox{MeV}/c^2$ bin at 0.6
$\hbox{GeV}/c^2$ to about 30 events at about 1.2 $\hbox{GeV}/c^2$.

\section{\boldmath Background Analysis}\label{bg-analysis}

Backgrounds are studied using the $\psip$
inclusive decay MC sample. By applying the same selection
criteria, it is found that for $\psip \to p \bar{n} \pi^-$ the
main backgrounds are from $\psip \to \gamma \chi_{cJ}, \chi_{cJ}
\to p \bar{n} \pi^- \pi^0 $, $\psip \to p \bar{n} \pi^- \pi^0$, and
$\psip \to \gamma \chi_{cJ}, \chi_{cJ} \to p \bar{n} \pi^-$. All
these modes have not been measured previously. Here we measure
these channels.

For these three background channels, in addition to the above
selection criteria, we also require:
\begin{enumerate}
\item For $\psip \to \gamma \chi_{cJ}, \chi_{cJ} \to p \bar{n}
  \pi^- \pi^0 $ events, we do a one-constraint (1C) kinematic fit to
  $\psip \to \gamma \gamma \gamma p \bar{n} \pi^-$ looping over all
  photon candidates.  The combination with the minimum $\chi^2$ is
  selected, and $Prob_{1C}>0.01$ is required. After the 1C fit, the
  direction of the $\bar{n}$ is determined. Similar requirements are
  imposed on $\psip \to \gamma \gamma p \bar{n} \pi^-$ and $ \psip \to
  \gamma p \bar{n} \pi^-$ final states in selecting $\psip \to p
  \bar{n} \pi^- \pi^0$ and $\psip \to \gamma \chi_{cJ}, \chi_{cJ} \to
  p \bar{n} \pi^-$.
\item The angle between the direction of $\bar{n}$ and one of the
  neutral clusters not already used in the kinematic fit should be less than
  ten degrees.
\end{enumerate}

\subsection{\boldmath $\psip \to \gamma \chi_{cJ}, \chi_{cJ} \to p \bar{n}
\pi^- \pi^0 $}

\begin{figure}[htbp]
\begin{center}
\epsfig{file=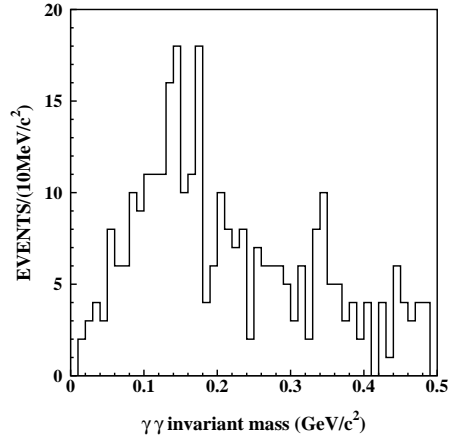,width=6.5cm} \caption{\label{mgg} The $\gamma
\gamma$ invariant mass distribution for selected $\psip \to \gamma p
\bar{n} \pi^- \gamma \gamma $ candidate events.}
\end{center}
\end{figure}

For this channel, we require four photon candidates (three of them
from the radiative and $\pi^0$ decays, the other from the
interaction between the antineutron and the detector material) and
the angle between the two photons from $\pi^0$ decays be greater
than eight degrees to remove the background from split-off fake
photons. In order to suppress the background from $\psip\to
\pi^0\pi^0 J/\psi, \jpsi \to p \bar{n}\pi^-$, the requirement $|m_{p
\bar{n}\pi^-}-3.097|>0.05$ $\hbox{GeV}/c^2$ is used. After applying
the above selection criteria, the $\gamma \gamma$ invariant mass
spectrum, shown in Fig.~\ref{mgg}, is obtained, and $|m_{\gamma
\gamma}-0.135|<0.03$ $\hbox{GeV}/c^2$ is required to select $\pi^0$
candidates.

\begin{figure}[htbp]
\begin{center}
\epsfig{file=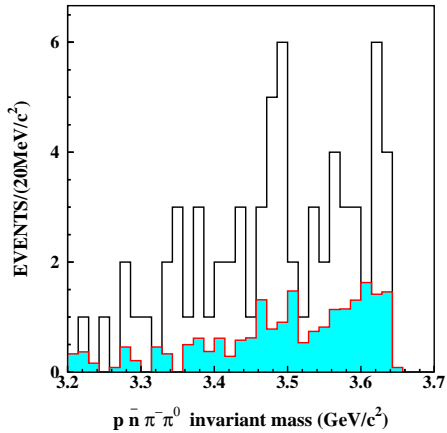,width=6.5cm} \caption{\label{gchicj-pnpim} The
$p \bar{n} \pi^- \gamma \gamma$ invariant mass distribution for
selected $\psip \to \gamma  p \bar{n} \pi^- \pi^0 $ candidate
events. The shaded histogram is the sum of $\psip\to \pi^0 \pi^0
\jpsi, \jpsi\to p \bar{n} \pi^-$ and $\psip\to p \bar{n} \pi^-
\pi^0$ backgrounds which have been normalized to data according to
their branching fractions.}
\end{center}
\end{figure}

The $p \bar{n} \pi^- \gamma \gamma$ invariant mass distribution for
the selected $\psip \to \gamma p \bar{n} \pi^- \pi^0 $ candidate
events is shown in Fig.~\ref{gchicj-pnpim}. The shaded histogram
shows the main backgrounds from $\psip\to \pi^0 \pi^0 \jpsi,
\jpsi\to p \bar{n} \pi^-$ and $\psip\to p \bar{n} \pi^- \pi^0$,
which have been normalized to data using the branching fractions of
$\jpsi\to p \bar{n} \pi^-$ and $\psip\to p \bar{n} \pi^- \pi^0$ from
Ref.~\cite{jxb} and our measurements (see
Sec.~\ref{pnpipi0-measure}). From Fig.~\ref{gchicj-pnpim}, no clear
$\chi_{cJ}$ signals are seen. Since the background in this channel
is very complicated, we set an upper limit for $\chi_{cJ}\to p
\bar{n} \pi^- \pi^0$ by subtracting the known backgrounds from the
total number of observed events and obtain $45\pm 9$ events for $p
\bar{n} \pi^- \pi^0$ invariant mass greater than 3.2
$\hbox{GeV}/c^2$. Assuming a Gaussian distribution, the upper limit
at the 90\% C. L. is 57. Using the MC simulated efficiency
$\epsilon_1$ of $(4.00\pm0.09)\%$, we obtain:
 \begin{equation*}
 \sum_{J=0}^2 B(\psip \to \gamma \chi_{cJ},\chi_{cJ} \to p \bar{n}
\pi^- \pi^0)
 <\frac {N^{up}_{\gamma p \bar{n}
\pi^- \pi^0}} {N_{\psip}\cdot \epsilon_1\cdot f_1\cdot f_3\cdot
(1-s)\cdot B(\pi^0\to\gamma\gamma)}
 =1.2\times 10^{-4},
\end{equation*}
at the $90\%$ C. L. Here $f_1$ and $f_3$ are efficiency correction
factors (see Sec.~\ref{systematic-error}, items 6 and 8) and $s$ is
the systematic error (see Table~\ref{sumerror}).

\subsection{\boldmath $\psip \to p \bar{n} \pi^- \pi^0 $}
\label{pnpipi0-measure}

\begin{figure}[htbp]
\begin{center}
\epsfig{file=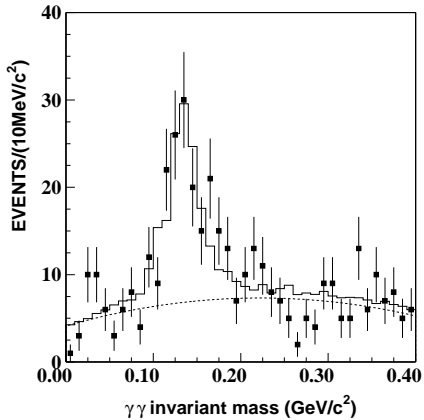,width=6.cm} \caption{\label{pnpipi0-fit}The
$\gamma \gamma$ invariant mass distribution of the selected $ \psip
\to \gamma \gamma p \bar{n} \pi^-$ candidate events. The squares
with error bars are data, and the histogram is the fit. }
\end{center}
\end{figure}

For this channel, we require the number of selected photon
candidates equals three (two of them come from $\pi^0$ decays and
the other from the interaction of the antineutron with the detector
material) and the angle between the two photons from $\pi^0$ decays
be greater than eight degrees to remove the background from
split-off fake photons. The $\gamma \gamma$ invariant mass
distribution after applying all the above selection criteria is
shown in Fig.~\ref{pnpipi0-fit}, where a clear $\pi^0$ signal can be
seen. From the exclusive MC simulation, it is seen that possible
backgrounds, except $\psip\to\gamma \chi_{cJ},\chi_{cJ}\to p \bar{n}
\pi^- \pi^0$, have no peak at the $\pi^0$ mass in the $\gamma\gamma$
invariant mass distributions, and therefore they will not contribute
to the number of $\pi^0$ events in fitting the $\gamma \gamma$
invariant mass distribution. For $\psip\to\gamma
\chi_{cJ},\chi_{cJ}\to p \bar{n} \pi^- \pi^0$ background, after
normalizing to data according to the branching fraction measured
above, the contribution of this background is so small that it can
be neglected.

By fitting the $\gamma\gamma$ invariant mass spectrum with a MC
determined signal shape and a 2nd order Legendre polynomial for background,
as shown in Fig.~\ref{pnpipi0-fit}, $135\pm 21$ $\psip \to p \bar{n}
\pi^- \pi^0$ candidate events are obtained. The statistical
significance is $8.1\sigma$, and
the detection efficiency $\epsilon_2$ for this decay mode is
$(3.21\pm0.08)\%$. We obtain:
\begin{equation*}
 B(\psip \to p \bar{n} \pi^-\pi^0)=
 \frac{N^{sig}_{p \bar{n} \pi^-\pi^0}}{N_{\psip}\cdot
 \epsilon_2\cdot f_1\cdot f_3 \cdot
 B(\pi^0\to\gamma\gamma)} =(3.18\pm 0.50)\times 10^{-4},
\end{equation*}
where $f_1$, $f_3$ are the efficiency correction factors, and the
error is statistical.

\subsection{\boldmath $\psip \to \gamma \chi_{cJ}, \chi_{cJ}
\to p \bar{n} \pi^-$}

\begin{figure}[htbp]
\begin{center}
\epsfig{file=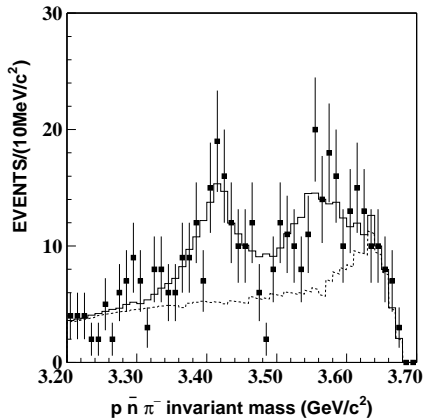,width=6.0cm} \caption{\label{cj-pnpim-fit} The
$p \bar{n} \pi^-$ invariant mass distribution of the selected $\psip
\to \gamma  p \bar{n} \pi^-$ candidate events. The squares with
error bars are data, the histogram is the fit, and the dashed curve
is the background shape from the fit. }
\end{center}
\end{figure}

For this channel, we require two photon candidates (one comes from
radiative decay and the other one from the interaction between the
antineutron and the detector material). After applying all the above
selection criteria, the $p \bar{n} \pi^-$ invariant mass
distribution, shown in Fig.~\ref{cj-pnpim-fit}, is obtained. Here
$\chi_{c0}$ and $\chi_{c2}$ signals can be seen, but the $\chi_{c1}$
signal is less significant. In fitting, we neglect the $\chi_{c1}$
signal, and $85\pm 18$ and $80 \pm 16$ events are obtained by
fitting the $p \bar{n} \pi^-$ invariant mass spectrum with MC
simulated $\chi_{c0}$ and $\chi_{c2}$ signal histograms and a
background shape. Here the background shape includes a 2nd order
Legendre function and a normalized histogram obtained from
background channel $\psip \to p \bar{n} \pi^-$ measured in this
work.

The statistical significance of $\chi_{c0}$ and  $\chi_{c2}$ are
both $4.2\sigma$, and the detection efficiencies for $\chi_{c0}$
($\epsilon_3$) and $\chi_{c2}$ ($\epsilon_4$) are $(5.77\pm
0.11)\%$ and $(6.14\pm 0.11)\%$, respectively. We obtain:
\begin{equation*}
 B(\psip \to \gamma\chi_{c0}, \chi_{c0} \to p \bar{n}
 \pi^-)=\frac{N^{sig}_{\chi_{c0} \to p \bar{n}
 \pi^-}}{N_{\psip}\cdot \epsilon_3 \cdot f_1 \cdot f_3}
 =(1.10\pm0.24)\times 10^{-4},
\end{equation*}
\begin{equation*}
 B(\psip \to \gamma \chi_{c2}, \chi_{c2}
\to p \bar{n} \pi^-)= \frac{N^{sig}_{\chi_{c2} \to p \bar{n}
\pi^-}}{N_{\psip}\cdot \epsilon_4 \cdot f_1\cdot f_3 }
=(0.97\pm0.20)\times 10^{-4},
\end{equation*}
where $f_1$ and  $f_3$ are the efficiency correction factors, and the
errors are statistical.

Since the neutron passes through the detector without interaction in
almost all cases, the background level will be higher and more
complicated than for the antineutron final state, so we do not try
to measure the charge conjugate modes of the above three channels.
In the following analysis, we assume the branching fractions for
these three channels equal to their charge conjugate modes.

\section{\boldmath Systematic Errors}\label{systematic-error}

Systematic errors in measuring the branching fractions mainly
originate from the MC statistics, the error matrix of the track
finding and the 1C kinematic fit, the vertex requirement, particle
identification, the photon efficiency, the $\alpha$ selection
criterion, the total number of $\psip$ events, and the fitting of
the signal.

\begin{enumerate}
\item The MDC tracking efficiency was measured using channels like
$J/\psi \to \Lambda \bar{\Lambda}$ and $\psip \to \pi^+\pi^-
J/\psi, J/\psi \to \mu^+ \mu^-$. It is found that the MC
simulation agrees with data within $(1-2)\%$ for each charged
track. Therefore, $4\%$ is taken as the systematic error for
events with two charged tracks.

\item The difference in $V_z$ between the positive and negative
charged tracks $|V_z^+ -V_z^-|$ is required to be less than 3~cm,
this corresponds to an about three-standard-deviation requirement.
The effect of it is checked with $\jpsi \to p \bar{n} \pi^- + c.c.$
candidate events, it is found that MC simulates data within 0.6\%.
This is taken as systematic error of this selection criterion.

\item  The photon detection efficiency was studied with different
methods using $J/\psi \to \pi^+\pi^-\pi^0$ events~\cite{simbes}, and
the difference between data and MC simulation is about $2\%$ for
each photon. We take $2\%$ per photon in the analysis.

\item Particle identification is used in selecting candidate
events, and we take $5\%$ as the systematic error~\cite{jxb}.

\item The systematic error from the 1C kinematic fit should be smaller
than for the 4C kinematic fit, since there are fewer constraints. Various
studies show that the uncertainty of the 4C kinematic fit is around
$4\%$~\cite{mall}, so here we conservatively take 4\% as the error from the 1C
kinematic fit.

\item The uncertainties of background shapes in $\psip \to p
\bar{n} \pi^-$, $\psip \to \bar{p} n \pi^+ $, $\psip \to p \bar{n}
\pi^- \pi^0 +c.c.$, $\psip \to \gamma \chi_{c0}, \chi_{c0} \to p
\bar{n} \pi^-$, and $\psip \to \gamma \chi_{c2}, \chi_{c2} \to p
\bar{n} \pi^-$ are estimated to be about $3.4\%$,
$4.0\%$, $12\%$, $14\%$, and $25\%$, respectively, by changing the
order of the background polynomial and the fitting range.

\item The effect of the requirement on the angle between the
nearest neutral cluster and the missing momentum direction of all
charged tracks is checked with $\jpsi \to p \bar{n} \pi^-
+c.c.$ candidate events, where the statistics are much higher and
the background is much lower. We require the same selection
criteria on these two channels as $\psip \to p \bar{n} \pi^-+c.c.$
except for the requirement on $\alpha$. By applying the requirement on
$\alpha$, we measure the efficiency of this selection criterion experimentally.

The efficiency difference for $\jpsi \to p \bar{n} \pi^-$ between
data and MC simulation is measured  to be $\frac{\epsilon_{DT}}
{\epsilon_{MC}}=(101.60\pm 0.53)\%$ . We take $f_1=1.02$ as the
efficiency correction factor for channels containing a $\bar{n}$.

The efficiency difference for $\jpsi \to \bar{p} n \pi^+$ is
measured  to be $\frac{\epsilon_{DT}} {\epsilon_{MC}} = (89.80\pm
0.49)\%$. We take $f_2=0.90$ as the efficiency correction factor for
channels containing a $n$. The big difference in the efficiency is
due to the fact that the simulation of the hadronic interaction of
the neutron with the detector material is not very reliable. The
systematic error on the efficiency associated with this requirement
is taken as 0.5\% .

\item The energy observed in the BSC associated with the positive
charged track is required to be less than 0.7~GeV. The effect of
this selection criterion is checked with $\jpsi \to  \bar{p} n
\pi^+$ candidate events, the efficiency difference between data
and MC simulation is measured to be
$\frac{\epsilon_{DT}}{\epsilon_{MC}}=(100.21\pm 0.27)\%$, and we
take 0.5\% as the systematic error on the efficiency associated
with this selection criterion.

\item In order to select $\psip \to \gamma \chi_{cJ}, \chi_{cJ}
\to p \bar{n} \pi^- \pi^0$, $\psip\to p \bar{n} \pi^- \pi^0$, and
$\psip \to \gamma \chi_{cJ}, \chi_{cJ} \to p \bar{n} \pi^-$
candidate events, four, three, and two photons are required,
respectively. The effect of these selection criteria are checked
with $\jpsi \to p \bar{n} \pi^-$ candidate events, and the
efficiency difference between data and MC simulation is measured
to be $\frac{\epsilon_{DT}}{\epsilon_{MC}}= (93.8\pm 2.2)\%$.
Because the MC simulates less fake photons than in data. We take
$f_3=0.94$ as the efficiency correction factor for three
background channels analyzed, and take 2.4\% as the systematic
error on the efficiencies associated with these three selection
criteria.

\end{enumerate}

Table~\ref{sumerror} lists all the systematic errors, and the
total systematic errors for $\psip \to p \bar{n} \pi^-$, $\psip
\to \bar{p} n \pi^+$, $\psip\to p \bar{n} \pi^- \pi^0$, $\psip\to
\gamma \chi_{c0}$, $\chi_{c0}\to p \bar{n} \pi^-$, and $\psip\to
\gamma \chi_{c2}$, $\chi_{c2}\to p \bar{n} \pi^-$ are 8.4\%,
8.6\%, 16\%, 17\%, and 27\%, respectively. The systematic error
for $\psip\to \gamma \chi_{cJ}$, $\chi_{cJ}\to p \bar{n} \pi^-
\pi^0$ is 11\%.

\begin{table*}[hbtp]
\begin{center}
\caption {Summary of systematic errors ($\%$), the errors common to
all modes are only listed once, and ``$\cdots$" means no
contribution.}
\begin{tabular}{l c c c c c c}
  \hline\hline
  Source &$p \bar{n} \pi^-$& $\bar{p} n \pi^+$ & $\gamma \chi_{cJ}\to \gamma p \bar{n} \pi^- \pi^0$ &
  $ p \bar{n} \pi^- \pi^0 $ &$ \gamma \chi_{c0}\to \gamma p \bar{n} \pi^-$ & $\gamma
\chi_{c2}\to \gamma p \bar{n} \pi^-$\\\hline
MC statistics       &  0.6  & 0.5 & 2.3 &2.3  & 2.0  & 1.8\\
1C kinematic fit    & $\cdots$   & $\cdots$ & 4.0 &  4.0  & 4.0 & 4.0     \\
Photon efficiency   &  $\cdots$  & $\cdots$ & 6.0  & 4.0   &  2.0    & 2.0\\
Fitting             & 3.4 & 4.0 & $\cdots$ & 12 &14  &25 \\
$E_+$              & $\cdots$ & 0.5 & $\cdots$ & $\cdots$ & $\cdots$  & $\cdots$ \\
$N_{\gamma}$        &$\cdots$& $\cdots$ &2.4 & 2.4 & 2.4 & 2.4 \\
Tracking error      & \multicolumn{6}{c}{4.0} \\
$|V_z^+ -V_z^-|$    & \multicolumn{6}{c}{0.6} \\
PID efficiency      & \multicolumn{6}{c}{5.0} \\
$\alpha$            & \multicolumn{6}{c}{0.5} \\
 $N_{\psip}$         & \multicolumn{6}{c}{4.0}
\\\hline
Sum                 & 8.4  &8.6 & 11 &16  & 17    & 27 \\
\hline\hline
\end{tabular}
\label{sumerror}
\end{center}
\end{table*}

\section{\boldmath Results and Discussion}

For the signal channel $\psip \to p \bar{n} \pi^- +c.c.$, the
branching fractions of the main background channels are given
above. From the inclusive and exclusive MC simulations, the
contributions of other background channels in the missing mass
distributions of $p \pi^-$ and $\bar{p} \pi^+$ show no peak.  By
fitting the missing mass distributions of $p \pi^-$ and $\bar{p}
\pi^+$ with signal histograms obtained from MC simulation,
normalized histograms of the MC simulated main background channels
measured in this analysis, and first order Legendre functions to
describe the other backgrounds, $921\pm 40$ and $914\pm 42$ events
are obtained. The fits are shown in Fig.~\ref{pnpi-data}. The
efficiency for $\psip \to p \bar{n} \pi^-$ is
$\epsilon_5=(26.48\pm0.14)$\%, and the branching fraction is:
\begin{equation*}
  B(\psip \to p \bar{n} \pi^-)= \frac{N^{sig}_{p \bar{n} \pi^-}}
  {N_{\psip}\cdot\epsilon_5\cdot
 f_1} =(2.45\pm0.11\pm0.21 )\times 10^{-4}.
\end{equation*}
The efficiency for $\psip \to \bar{p}  n \pi^+ $ is
$\epsilon_6=(28.83\pm0.15)$\%, and the branching fraction is:
\begin{equation*}
 B(\psip \to  \bar{p} n \pi^+)=\frac{N^{sig}_{\bar{p}  n \pi^+}}
 {N_{\psip}\cdot\epsilon_6\cdot
 f_2}=(2.52\pm0.12\pm 0.22 )\times 10^{-4}.
\end{equation*}
Here $f_1, f_2$ are the correction factors to the efficiencies, and
the first errors are statistical and the second ones are
systematic.

\begin{figure}[htbp]
\centerline{\psfig{file=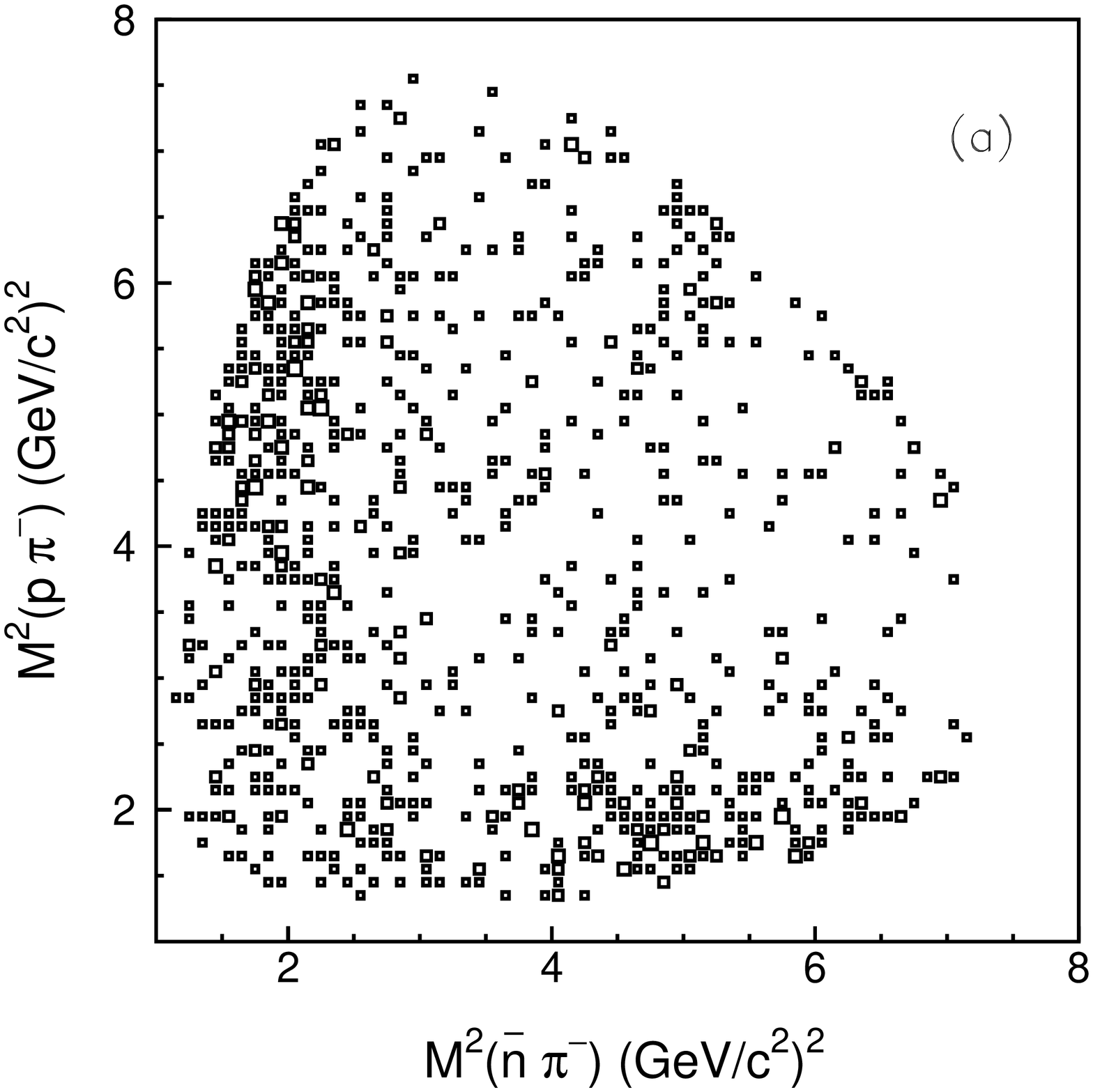,width=6cm}
            \psfig{file=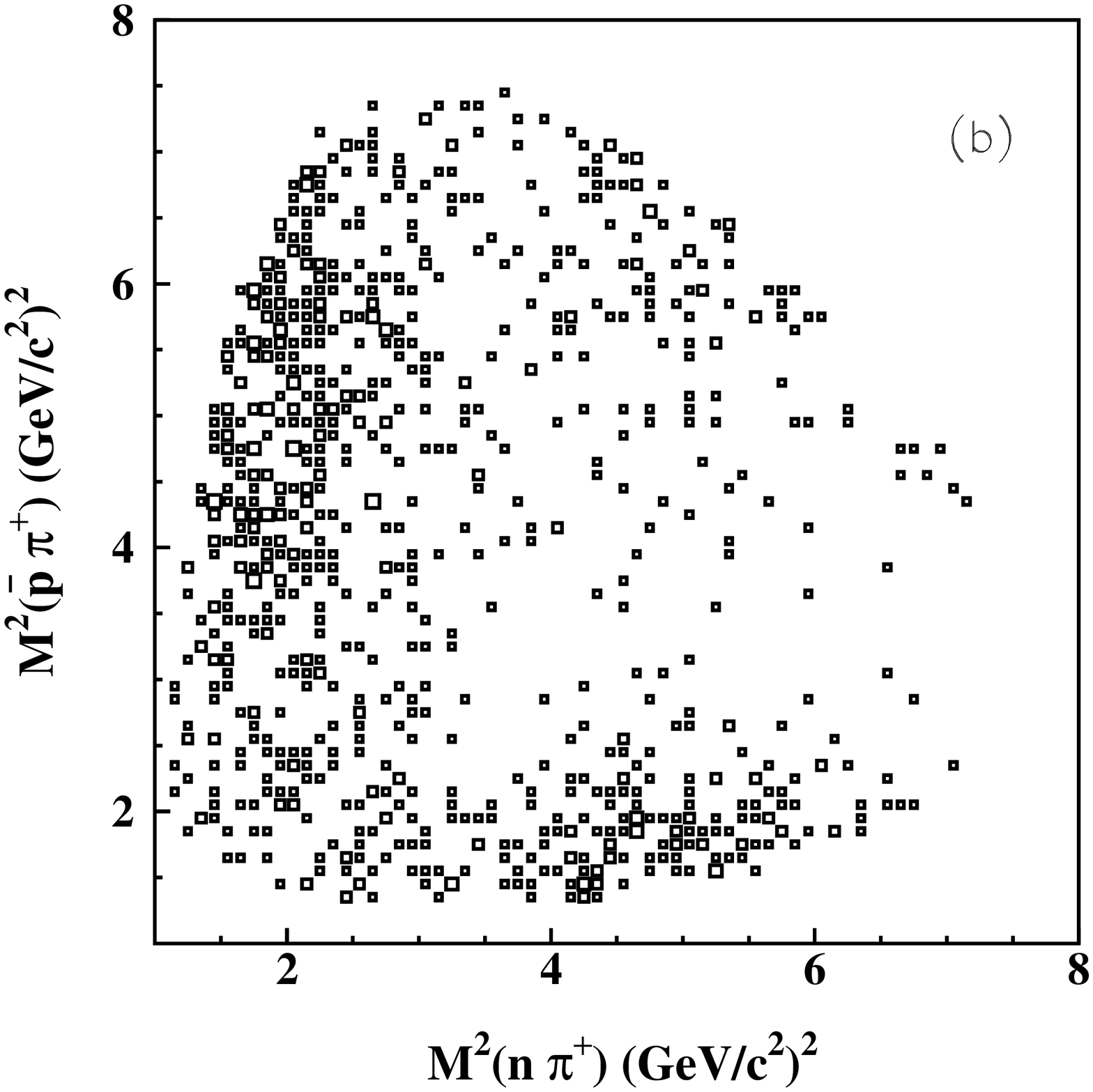,width=6cm}}
 \caption{\label{dalitz} Dalitz plots for (a) $\psip\to p \bar{n}
\pi^-$ and (b) $\psip \to \bar{p} n \pi^+$. Here we require
$|m^{recoil}_{p \pi}-0.938|<0.1$ $\hbox{GeV}/c^2$.}
\end{figure}

Taking events with the missing mass within $\pm 0.1$~$\hbox{GeV}/c^2$
around the neutron mass, we get 851 and 849 for $\psip \to p \bar{n}
\pi^-$ and $\psip \to \bar{p} n \pi^+$ events, respectively. The
Dalitz plots for these two channels, which are similar, are shown
in Fig.~\ref{dalitz}. The asymmetry between $p \pi$ and $n \pi$ is
partly due to the difference in detection efficiency and may be partly
due to isospin symmetry breaking effects from the electromagnetic
interaction~\cite{jpsidecay}. From the plots, the contribution of the
$N^{\ast}$ states at around $1.4-1.5$ $\hbox{GeV}/c^2$ can be seen,
and there is a possible vertical band at around $m^{2}=4.75$
$(\hbox{GeV}/c^2)^2$. A similar band in the horizontal direction is even
less clear since the events on the right side, where the recoil proton
(antiproton) has low momentum and cannot be detected well by the
detector, have low efficiency.

\begin{figure}[htbp]
\centerline{\psfig{file=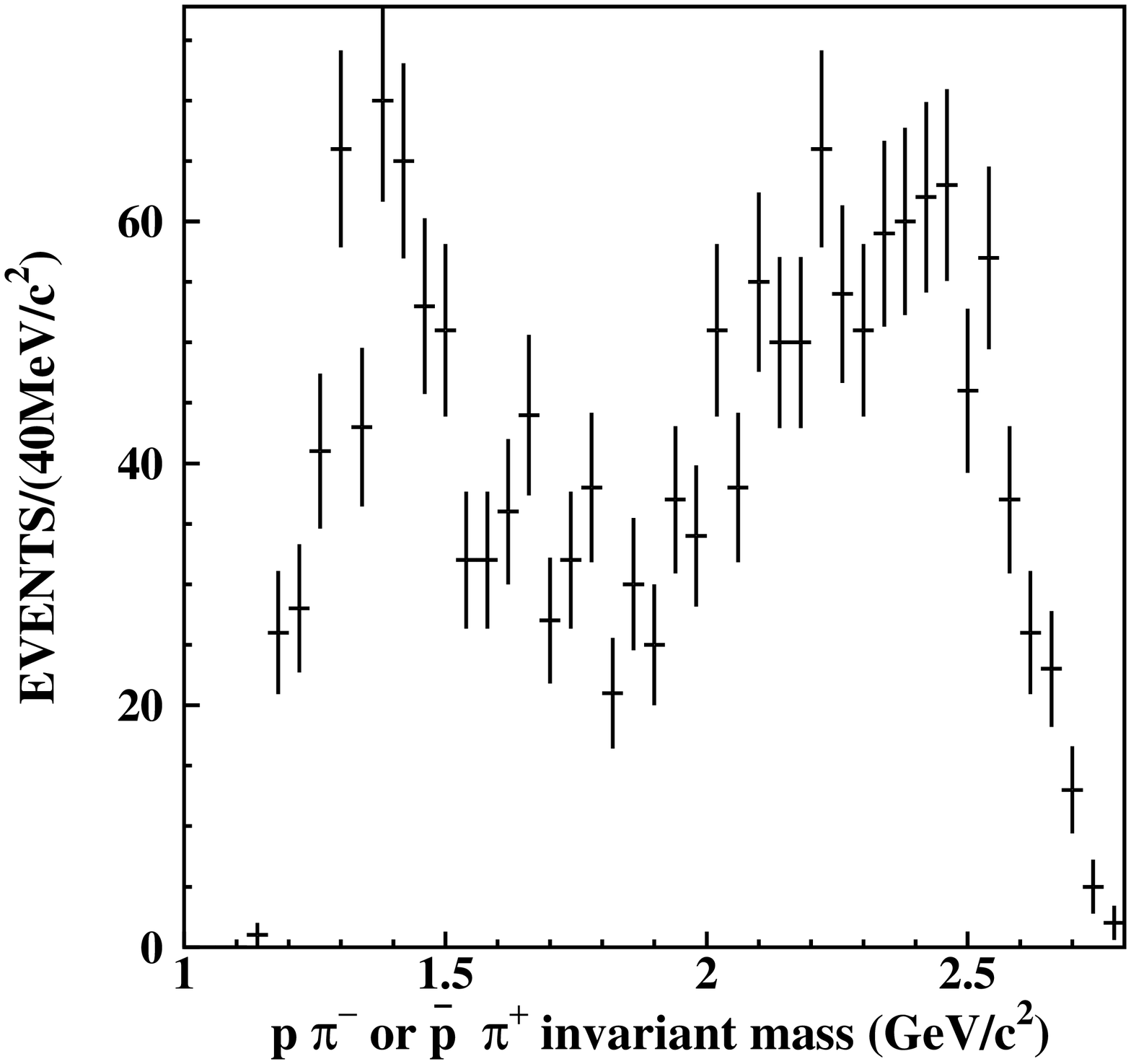,width=6cm}
            \psfig{file=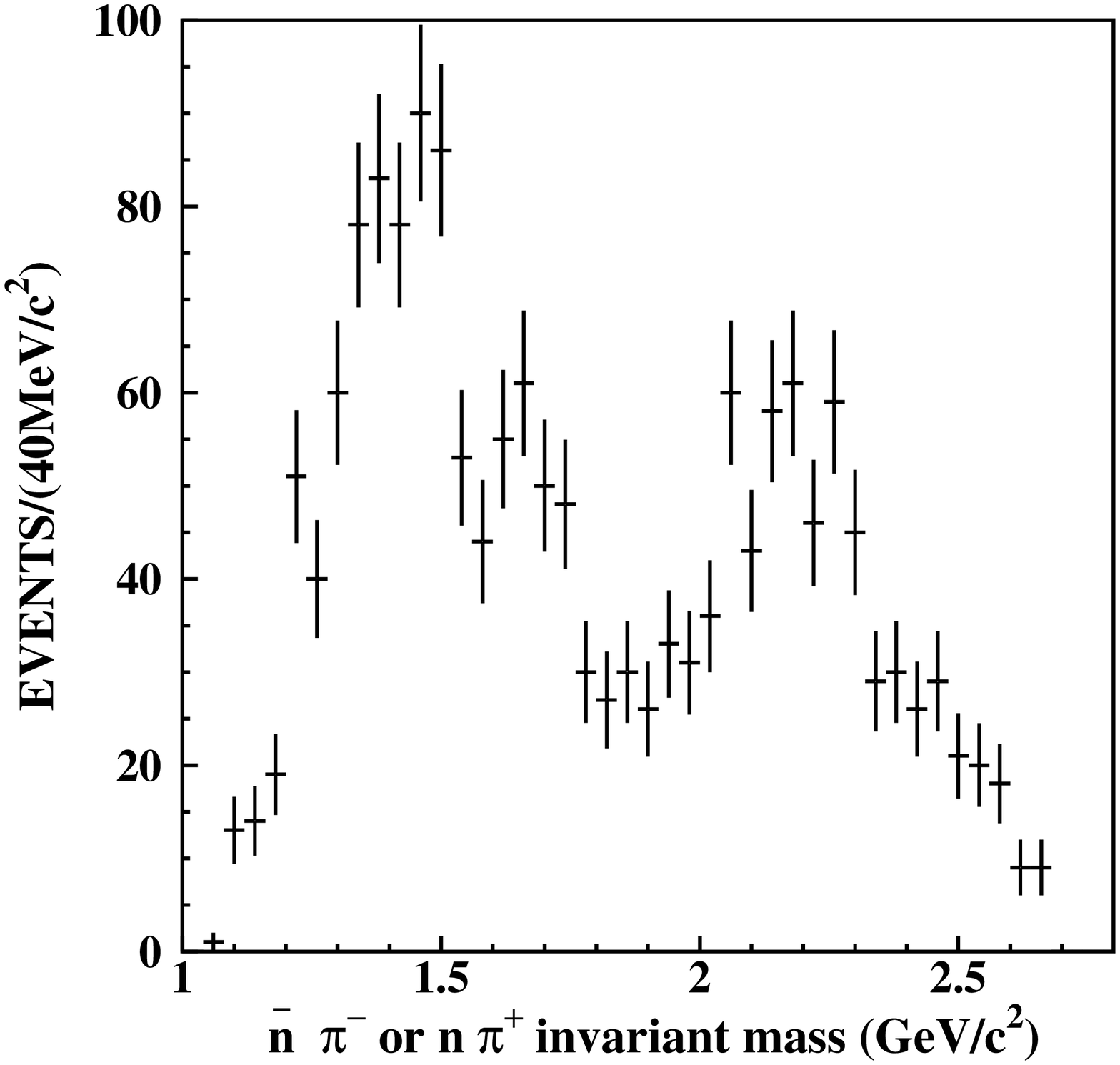,width=6cm}}
\caption{\label{am-before} $p \pi^-$ (or $\bar{p} \pi^+$) and
$\bar{n} \pi^-$ (or $n \pi^+$) invariant mass distributions for
$\psip \to p \bar{n} \pi^-+c.c.$}
\end{figure}

\begin{figure}[htbp]
\centerline{\psfig{file=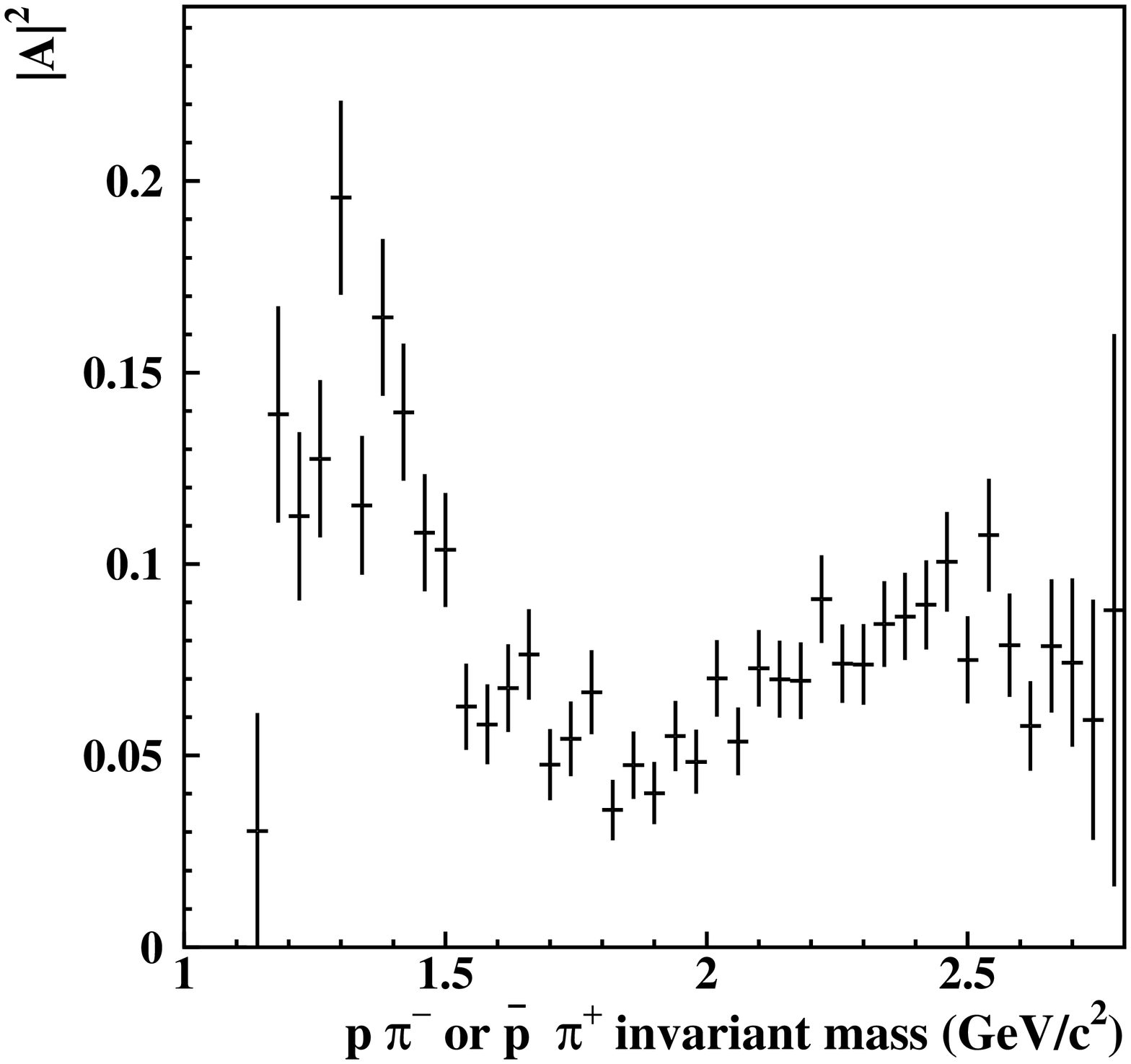,width=6cm}
            \psfig{file=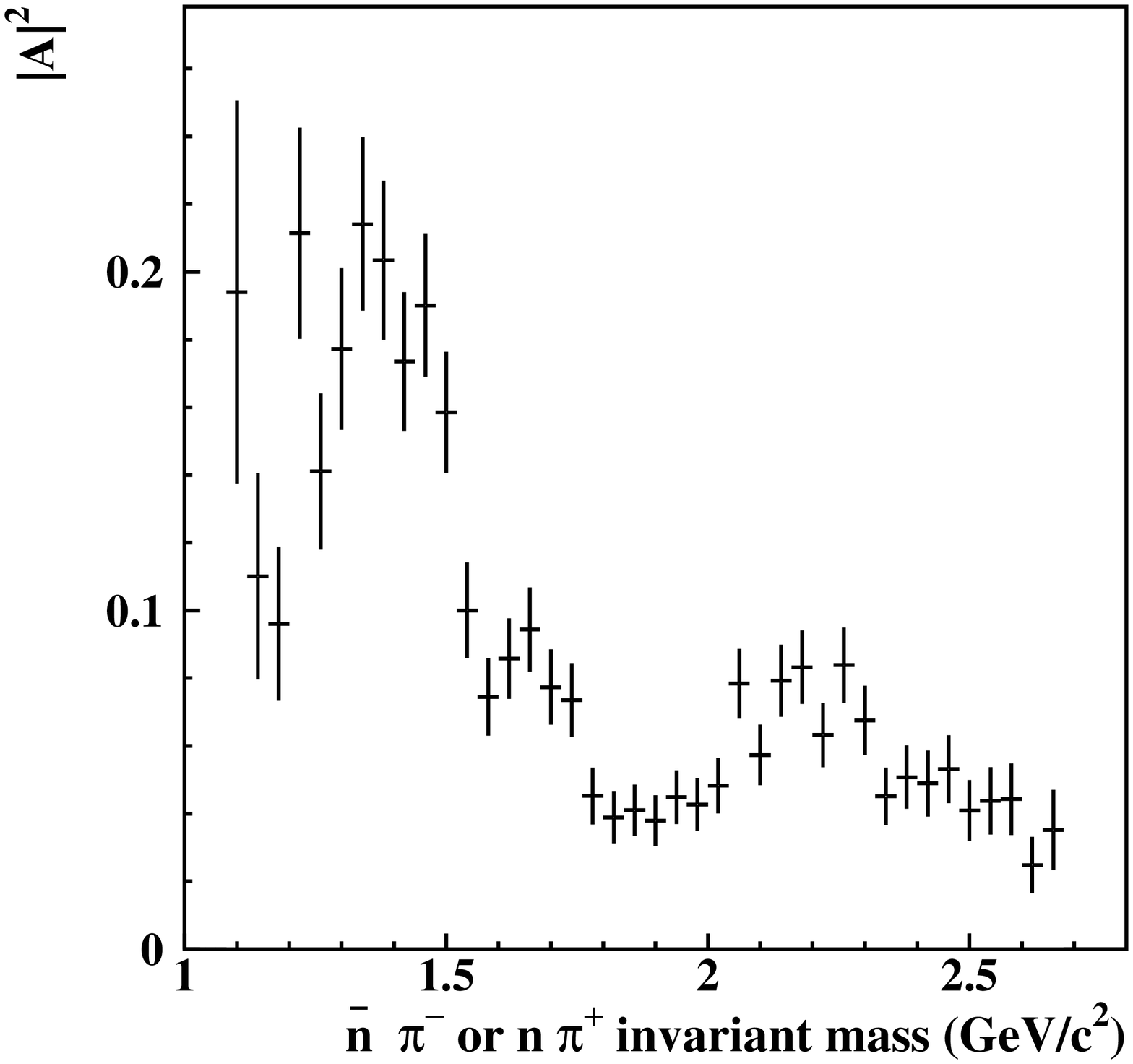,width=6cm}}
\caption{\label{am} Data corrected by MC simulated efficiency and
phase space versus $p \pi^-$ (or $\bar{p} \pi^+$) and $\bar{n}
\pi^-$ (or $n \pi^+$) invariant mass for $\psip \to p \bar{n}
\pi^-+c.c.$ candidate events.}
\end{figure}

Figure~\ref{am-before} shows the $p \pi^-$ (or $\bar{p} \pi^+$) and
$\bar{n} \pi^-$ (or $n \pi^+$) invariant mass distributions for
$\psip \to p \bar{n} \pi^-+c.c.$ In order to investigate the
behavior of the amplitude squared as a function of invariant
mass, the invariant mass distributions are divided by phase
space  and corrected by the mass dependent efficiency. The
results are shown in Fig.~\ref{am}. There is a large accumulation of events
below 1.5 $\hbox{GeV}/c^2$, which may be due to $N^{\ast}(1440)$,
$N^{\ast}(1520)$, $N^{\ast}(1535)$, etc. The cluster of events above
2 $\hbox{GeV}/c^2$ is partly due to the reflection of the
$N^{\ast}(1440)$ etc., and partly may due to high mass $N^{\ast}$ states,
for example $N^{\ast}(2190)$, $N^{\ast}(2220)$, $N^{\ast}(2250)$. No
clear $N^{\ast}(2065)$ peak is observed in the plots, although we
can not rule out its existence. Partial wave analysis is necessary
to obtain more information about the $N^{\ast}$ components in the data.
However, here the statistics are not large enough for such an
analysis.

Using the $\jpsi$ decay results in Refs.~\cite{pnpi1,pnpi2,pnpi3},
we obtain:
$$Q_{p \bar{n} \pi^-}=\frac{B(\psip \to p \bar{n} \pi^-)}
{B(\jpsi\to p \bar{n} \pi^-)}= (12.0\pm1.5)\%,  $$
$$Q_{\bar{p} n \pi^+}=\frac{B(\psip\to \bar{p} n \pi^+)}
{B(\jpsi\to\bar{p} n \pi^+)}= (12.9\pm1.7)\%, $$ which agree with
the $12\%$ rule within one standard deviation.

Using the branching fraction of $\psip\to p \bar{p} \pi^0$ from
Ref.~\cite{xinbo}, the ratio of $B(\psip\to p
\bar{p}\pi^0):B(\psip\to p \bar{n}\pi^-):B(\psip\to \bar{p} n
\pi^+)$ is measured to be $1:(1.86\pm0.27):(1.91\pm0.27)$. In
calculating this ratio, the common systematic errors between them
have been removed. This ratio is consistent with the ratio
$1:2:2$ predicted by isospin symmetry.

\section{\boldmath Summary}

Using 14 million $\psi(2S)$ events, the branching fractions of
$\psip \to p \bar{n} \pi^-$, $\psip \to \bar{p} n \pi^+$, $\psip
\to p \bar{n} \pi^- \pi^0$, $\psip \to \gamma \chi_{c0}, \chi_{c0}
\to p \bar{n} \pi^-$, and $\psip \to \gamma \chi_{c2}, \chi_{c2}
\to p \bar{n} \pi^-$ are measured to be:
$$
 B(\psip \to p \bar{n} \pi^-)=(2.45\pm0.11\pm0.21 )\times 10^{-4},
$$
$$
B(\psip \to  \bar{p} n \pi^+)=(2.52\pm0.12\pm 0.22 )\times 10^{-4},
$$
$$
 B(\psip \to p \bar{n} \pi^-\pi^0)=(3.18\pm0.50\pm0.50)\times 10^{-4},
$$
$$
 B(\psip \to \gamma\chi_{c0}, \chi_{c0} \to p \bar{n}
 \pi^-)=(1.10\pm0.24\pm 0.18)\times 10^{-4},
$$
$$
B(\psip \to \gamma \chi_{c2}, \chi_{c2} \to p \bar{n}
\pi^-)=(0.97\pm0.20\pm0.26)\times 10^{-4},
$$
where the first errors are statistical and the second are
systematic. The upper limit of $\sum_{J=0}^2 B(\psip \to \gamma
\chi_{cJ}, \chi_{cJ} \to p \bar{n} \pi^- \pi^0)$ is estimated to
be $1.2\times 10^{-4}$ at the 90\%~C.~L.

\acknowledgments

The BES collaboration thanks the staff of BEPC for their hard
efforts. This work is supported in part by the National Natural
Science Foundation of China under contracts Nos. 10491300, 10225524,
10225525, 10425523, the Chinese Academy of Sciences under contract
No. KJ 95T-03, the 100 Talents Program of CAS under Contract Nos.
U-11, U-24, U-25, and the Knowledge Innovation Project of CAS under
Contract Nos. U-602, U-34 (IHEP), the National Natural Science
Foundation of China under Contract No. 10225522 (Tsinghua
University), and the Department of Energy under Contract No.
DE-FG02-04ER41291 (U Hawaii).


\begin{thebibliography}{**}
\bibitem{T.Appelquist} T.~Appelquist and H.~D.~Politzer,
Phys. Rev. Lett. {\bf 34}, 43 (1975); A.~De R\'ujula and
S.~L.~Glashow, Phys. Rev. Lett. {\bf 34}, 46 (1975).
\bibitem{Mark-II} M.~E.~B.~Franklin {\itshape et
al}. (Mark-II Collaboration), Phys. Rev. Lett. {\bf 51}, 963
(1983).
\bibitem{J.L.Rosner}J.~L.~Rosner, Phys. Rev. D {\bf 64}, 094002 (2001).
\bibitem{P.Wang}P.~Wang, C.~Z.~Yuan and X.~H.~Mo, Phys. Rev. D {\bf 70},
114014 (2004).
\bibitem{pro} Proceedings of the Workshop on the Physics of Excited
Nucleons, (NSTAR2002) eds. S.~A.~Dytman and E.~S.~Swanson, World
Scientific, 2003; (NSTAR2001) eds. D.~Drechsel and L.~Tiator, World
Scientific, 2001.
\bibitem{M.Ripani}M.~Ripani {\itshape et al}., Phys. Rev. Lett. {\bf 91},
022002 (2003); S.~Stepanyan {\itshape et al}., Phys. Rev. Lett.
{\bf 91}, 252001 (2003).
\bibitem{T.Nakano} T.~Nakano {\itshape et al}., Phys. Rev.
Lett. {\bf 91}, 012002 (2003).
\bibitem{pdg} S.~Eidelman {\itshape et al.} (Particle Data Group), Phys.
Lett. B {\bf 592}, 1 (2004).
\bibitem{S.Capstick} S.~Capstick and W.~Roberts, Prog. Part. Nucl.
Phys. {\bf 45}, S241 (2000).
\bibitem{zou} B.~S.~Zou, Nucl. Phys. A {\bf 675}, 167C (2000).
\bibitem{jxb} M.~Ablikim {\itshape et al}. (BES Collaboration),
hep-ex/0405030.
\bibitem{xinbo}  M.~Ablikim {\itshape et al}. (BES Collaboration),
Phys. Rev. D {\bf 71}, 072006 (2005).
\bibitem{detector} J. Z. Bai {\itshape et al.} (BES Collaboration),
Nucl. Instrum. Methods Phys. Res., Sect. A {\bf 458}, 627 (2001).
\bibitem{simbes} M.~Ablikim {\itshape et al.} (BES Collaboration),
Nucl. Instrum. Methods Phys. Res., Sect. A {\bf 552}, 344 (2005).
\bibitem{X.H.Mo}X.~H.~Mo {\itshape et al}., High Energy Phys. Nucl. Phys. {\bf 28}, 455
(2004). [hep-ex/0407055].
\bibitem{chenjc} J.~C.~Chen, G.~S.~Huang, X.~R.~Qi, D.~H.~Zhang and Y.~S.~Zhu, Phys. Rev. D {\bf 62}, 034003 (2000).
\bibitem{mall} M.~Ablikim {\itshape et al}. (BES Collaboration), Phys. Rev. D {\bf 70}, 112007 (2004).
\bibitem{jpsidecay} L.~K$\ddot{\hbox{o}}$pke and N.~Wermes, Phys. Rept. {\bf 174},
67 (1989).
\bibitem{pnpi1} M.~W.~Eaton {\itshape et al}., Phys. Rev. D {\bf 29}, 804 (1984).
\bibitem{pnpi2} H.~J.~Besch {\itshape et al}., Z.~Phys. C {\bf 8}, 1 (1981).
\bibitem{pnpi3} I.~Peruzzi, M.~Piccolo {\itshape et al}., Phys. Rev. D {\bf 17}, 2901 (1978).
\end{thebibliography}
\end{document}